\documentclass{ws-ijmpe}
\usepackage{graphicx}
\usepackage{epsfig}
\usepackage{amssymb}
\begin{document}

\catchline{}{}{}{}{}

\title{Effects of active-sterile neutrino mixing during primordial nucleosynthesis}
\author{Osvaldo Civitarese}
\address{\it{Department of Physics, University of La Plata \\49 y 115. c.c. 67 (1900), La Plata, Argentina}\\ osvaldo.civitarese@fisica.unlp.edu.ar}
\author{Mercedes Elisa  Mosquera}
\address{\it{Facultad de Ciencias Astron\'{o}micas y Geof\'{\i}sicas,  Universidad Nacional de La Plata\\ Paseo del Bosque, (1900) La Plata, Argentina.}\\ mmosquera@fcaglp.unlp.edu.ar}
\author{Mar\'{\i}a Manuela S\'aez}
\address{\it{Facultad de Ciencias Astron\'{o}micas y Geof\'{\i}sicas,  Universidad Nacional de La Plata\\ Paseo del Bosque, (1900) La Plata, Argentina.}}

\maketitle
\begin{history}
\received{(received date)} \revised{(revised date)}
\end{history}
\date{Received: date / Revised version: date}
\begin{abstract}
In the present work, we discuss the effects of the inclusion of
sterile-active neutrino oscillations during the production of
primordial light-nuclei. We assume that the sterile neutrino
mass-eigenstate might oscillate with the two lightest active
neutrino mass-eigenstates, with mixing angles $\phi_1$ and $\phi_2$.
We also allow a constant renormalization (represented by a parameter
$(\zeta)$) of the sterile neutrino occupation factor. Taking $\zeta$
and the mixing angles as free parameters, we have computed
distribution functions of active and sterile neutrinos and
primordial abundances. Using observable data we set constrains in
the free parameters of the model. It is found that the data on
primordial abundances are consistent with small mixing angles and
with a value of $\zeta$ smaller than $0.65$ at $3\sigma$ level.
\end{abstract}

\maketitle
\section{Introduction}
\label{intro}

The $^7$Li problem comes from the comparison between the
observational data and the theoretical results for the primordial
abundance of lithium when the value of the baryon density, obtained
from the analysis of the data from the Wilkinson Microwave
Anisotropy Probe (WMAP) and from Planck \cite{hinshaw13,planck13},
is used. There exist several attempts to solve this problem, such
as a better understanding of turbulent transport in the radiative
zone of stars \cite{richard05}, the existence of a stellar lithium
depletion \cite{melendez10,lind10}, and the inclusion of resonances
in reaction rates activated in the BBN decay path
\cite{kirsebom11,broggini12,cm12}. Despite the efforts, the problem
persists and the observed abundance of primordial lithium is smaller
than the predicted BBN abundance.

Neutrino flavor oscillations have been observed
\cite{chooz09,sage09,sno01,sk98,kamland03,gno05,k2k06,borexino08}
and described in terms of the mixing between three active neutrino
mass eigenstates, which constitutes the basis for each flavor. The
results of the Liquid Scintillator Neutrino Detector (LSND)
\cite{aguilar01} and the Mini Booster Neutrino Experiment
(MiniBooNE) \cite{aguilar07}, indicate the occurrence of some anomalies which may be
interpreted as possible signals of extra neutrino species.

The effects of active-sterile neutrino oscillations during
primordial nucleosynthesis have been analyzed by several authors
\cite{smith06,kishimoto06,keranen03,abazajian05}. In particular, in
Refs. \cite{bell99,dibari02,foot96} the effect of a neutrino
asymmetry was studied. Other authors
\cite{hannestad12,tamborra13,jacques13} have studied consequences of
assuming full and/or partial thermalization of the sterile neutrino distribution
during BBN. The active-sterile neutrino oscillations affect the
primordial abundances of light nuclei through the modification of
the beta decay rates.

In previous works we have focussed on the effect of active-sterile
neutrino oscillations during the epoch of light nuclei production.
In Ref. \cite{civitarese08a,civitarese08b,mosquera11} we have
discussed the effects of active-sterile neutrino oscillations in
different n+1 schemes(n active neutrinos, one sterile neutrino). In
Ref. \cite{mosquera14} we have performed the analysis in the 3 + 1
scheme, with a variable normalization of the sterile neutrino
sector, and we were able to set constrains on the mixing parameters
from the comparison between theoretical results and observable data.

In this work, we continue with our search for signals of sterile
neutrinos in BBN processes, by including one sterile neutrino which
can oscillate with two of the lightest active mass-eigenstates, and
by taking a variable normalization constant $(\zeta)$ for the
occupation factor of the sterile neutrino. We have solved quantum
kinetic equations (QKE) numerically in order to obtain the
distribution functions of active and sterile neutrinos. We have
solved these equations for two different cases: i) considering that
the neutrino-antineutrino interaction is null $(C=0)$; and, ii) with
$C\neq 0$, being $C$ the strength of the interaction, as it is explained in the next section.
 Using the available data on primordial abundances
(observational data) we set constrains on the free parameters of our
model, namely $\zeta$ and the two mixing angles.

This paper is organized as follows. In Section \ref{formalismo} we
describe the formalism to include active-sterile neutrino
oscillations in the production of light nuclei. In Section
\ref{resultados} we present and discuss the results of the
calculation of primordial abundances. We have extracted allowed
values of $\zeta$ and the mixing angles by performing a statistical
analysis of the calculated abundances. Finally, the conclusions are
drawn in Section \ref{conclusion}.

\section{Formalism}
\label{formalismo}

The inclusion of a new kind of neutrino, the sterile neutrino,
affects the neutrino density  of active neutrinos due to
active-sterile neutrino oscillation. In the 3+1 scheme there exists
three mixing angles between the sterile neutrino and active
neutrinos. In this work, we assume that the mixing between the
sterile neutrino and the heaviest neutrino mass eigenstate is null,
therefore we have considered two mixing angles, $\phi_1$ and
$\phi_2$.

In order to compute the statistical occupation factors of neutrinos
of a given flavor we follow the formalism of Ref. \cite{bell99}. One
can write neutrino densities as functions of a set of parameters
$P_i$, which depend on the neutrino energy and on the temperature,
then
\begin{eqnarray}
n_{\nu_e}&=&\frac{1}{2}P_0\left(1+P_3+\frac{1}{\sqrt{3}}P_8\right)n^{eq}\nonumber \\
n_{\nu_\mu}&=&\frac{1}{2}P_0\left(1-\frac{2}{\sqrt{3}}P_8\right)n^{eq}\nonumber \\
n_{\nu_s}&=&\frac{1}{2}P_0\left(1-P_3+\frac{1}{\sqrt{3}}P_8\right)n^{eq} ,
\end{eqnarray}
where $n_{\nu_e}$, $n_{\nu_\mu}$ and $n_{\nu_s}$ are 
occupation factors for the electron, muon and sterile neutrinos,
respectively, and $n^{eq}$ is the Fermi-Dirac distribution function for each type of particles.
The QKE for those parameters can be written as
\cite{mckellar94}
\begin{eqnarray}
\frac{d\textbf{P}}{dt}&=&\textbf{V}\times\textbf{P}-
D\left(P_1\hat{x_1}+P_2\hat{x_2}+P_6\hat{x_6}+P_7\hat{x_7}\right)  -D'\left(P_4\hat{x_4}+P_5\hat{x_5}\right)\nonumber \\
&&-C \left(\bar{P_4}\hat{x_4}-\bar{P_5}\hat{x_4}\right) +\frac{2}{3}\left[\left(\frac{3}{2}-P_3\right) \frac{R_e}{P_0}-P_3\frac{R_\mu}{P_0}\right]\hat{x_3}\nonumber \\
&&-\left(\frac{P_1}{P_0}\hat{x_1}+\frac{P_2}{P_0}\hat{x_2} +\frac{P_4}{P_0}\hat{x_4}+\frac{P_5}{P_0}\hat{x_5}  +\frac{P_6}{P_0}\hat{x_6} +\frac{P_7}{P_0}\hat{x_7}\right)\frac{d P_0}{dt}\nonumber \\
&&+\frac{2}{3}\left[\left(\frac{\sqrt{3}}{2}-P_8 \right)\frac{R_e}{P_0}-(\sqrt{3}+P_8)\frac{R_\mu}{P_0} \right]\hat{x_8}\nonumber \\
&&+\left(-P_6 \, {\rm Re} (H)-P_7 \, {\rm Im} (H)\right)\hat{x_1} +\left(-P_6 \, {\rm Im} (H)+P_7 \, {\rm Re} (H)\right)\hat{x_2}\nonumber \\
&&+\left(-P_1 \, {\rm Re} (H)-P_2 \, {\rm Im} (H)\right)\hat{x_6} +\left(-P_1 \, {\rm Im} (H)+P_2 \, {\rm Re} (H)\right)\hat{x_7}\nonumber \\
\frac{dP_0}{dt}&=&\frac{2}{3}\left(R_e+R_\mu\right) .
\label{QRE}
\end{eqnarray}
In the previous equation, $\textbf{P}$ is the vector with components
$P_i$, and $\textbf{V}$ is the effective potential
\begin{eqnarray}
\textbf{V}&=&2 \, {\rm Re} (E^{es})\hat{x_1} - 2 \, {\rm Im} (E^{es}) \hat{x_2} + \left(E^{ee}-E^{ss}\right) \hat{x_3} +2\, {\rm Re} (E^{e\mu}) \hat{x_4}- 2\, {\rm Im} (E^{e\mu}) \hat{x_5} \nonumber \\ 
&&+ 2\, {\rm Re} (E^{s\mu}) \hat{x_6} -2 \, {\rm Im} (E^{s\mu}) \hat{x_7} + \frac{1}{\sqrt{3}} \left(E^{ee}+E^{ss}-2E^{\mu\mu}\right) \hat{x_8} ,
\end{eqnarray}
where
\begin{eqnarray}
E^{\alpha\beta}=\left[\frac{1}{2p}U {\rm diag}
\left(m_1^2,m_2^2,m_3^2\right) U^{\dag}
\right]^{\alpha\beta}+V^{\alpha\beta} ,
\end{eqnarray}
$m_1$, $m_2$ and $m_3$ are the masses of the mass
eigenstates, $p$ is the neutrino momentum, and $U$ is the mixing
matrix
\begin{eqnarray}
U&=&\left(
\begin{array}{ccc}
c_1 c-s_1 s_2 s &s_1 c_2 & c_1 s + s_1 s_2 c \\
-s_1 c- c_1 s_2 s & c_1 c_2 & s_1 s+c_1 s_2 c \\
-c_2 s &-s_2 &c_2 c
\end{array}
\right) .
\end{eqnarray}
In this notation $s_i$ stands for $\sin \phi_i$, $c_i= \cos \phi_i$,
$s=\sin \theta$ and $c=\cos \theta$, where $\theta$ is the mixing
angle between the two active-neutrino mass eigenstates. The diagonal
terms of the neutrino interaction are written as
\begin{eqnarray}
V^{\alpha\alpha}&=&\frac{4 \zeta(3)
\sqrt{2} G_F T^3}{2 \pi^2 }\left[L^{\alpha} +A_\alpha\frac{T p}{M_W^2 }\right] ,
\end{eqnarray}
where $G_F$ is the Fermi constant, $T$ stands for the temperatures,
$\zeta (3)$ is a Reimann zeta-function, $M_W$ is the W-boson mass
and $L^{\alpha}$ is the lepton asymmetry. The values for the
constants $A_\alpha$ are $A_e\simeq 17$ and $A_{\mu, \, \tau}\simeq
4.9$ \cite{bell99}. The non-diagonal terms of the potential are
neglected, as well as the lepton asymmetry ($L^{\alpha}=0$).

The quantities $D$ and $D'$ of Eq.(\ref{QRE}) are damping parameters
\begin{eqnarray}
D&=&\frac{1}{2}G_F^{2}T^5 y_{e}\frac{p}{\left<p_0\right>}\nonumber \\
D'&=&\frac{1}{2} G_F^{2}T^5 y_{\mu}\frac{p}{\left<p_0\right>} ,
\end{eqnarray}
where $y_{e}=4$, $y_{\mu}=2.9$, and $\left<p_0\right>$ is the
averaged momentum for relativistic Fermi-Dirac distribution with
zero chemical potential \cite{bell99}. The parameter $C$ is the
strength of the coupling between the neutrino and antineutrino
density matrices, and it can be written as
\begin{eqnarray}
C&=&1.8 G_F^2 T^5 .
\end{eqnarray}
In Eq.(\ref{QRE}) $R_\alpha$ are re-population functions, and $H$ are exchange factors \cite{bell99} which are smaller
than the damping functions $D$ and $D'$ \cite{bell99}. To
solve these sixteen coupled differential equations (the unknowns are
the eight factors $P_i$ for neutrinos and the eight factors for
anti-neutrinos) we have assumed that the factor $P_0$ is
constant (due to the fact that the re-population factors are small),
and we have neglected the function $H$ and the non-diagonal terms on
the neutrino potential, as said before. We have performed the
calculations for two different cases: i) by considering that the
coupling between neutrino and anti-neutrino density matrices is null
($C=0$), and ii) taking $C \ne 0$.

The initial condition was set at $T_0=3$ MeV, and we have assumed
that the active neutrinos have standard Fermi-Dirac distributions at
that temperature. For the sterile neutrino we have considered two
different situations for its occupation factor, namely: i) a null
occupation factor; and,  ii) a Fermi-Dirac distribution multiplied
by a constant factor $\zeta$ \cite{ichikawa08,acero09} which can
vary from $0$ to $1$.

With these parameters and approximations, we have calculated
primordial abundances by using a modified version of the Kawano's
code \cite{kawano88,kawano92} (see Ref. \cite{mosquera11} for more
details on the modifications to the standard code). In the
calculations, the value of the square mass-difference (between the lightest active neutrino mass and the sterile neutrino mass) was fixed at $1$ eV$^2$ \cite{kopp13,archidiacono13}.

\section{Results}
\label{resultados}

The value of the active-neutrino mixing, namely the angle $\theta$
and the square-mass difference, have been determined from SNO, SK,
GNO, CHOOZ DAYA BAY, RENO and DOUBLE CHOOZ experiments
\cite{sno02,k2k06,gno05,chooz09,dayabay,reno,doublechooz}. The
baryon density was fixed at the value determined by WMAP
collaboration \cite{hinshaw13}.

The observational data for deuterium (D) have been extracted from
Refs.
\cite{noterdaeme12,pettini08,balashev10,pettini12,cooke14,fumagalli11,omeara06}.
We use the data from Refs.
\cite{villanova12,peimbert07,izotov10,aver12,villanova09,aver13,izotov13,aver10}
for $^4$He, and for $^7$Li we have considered the data given by
Refs. \cite{monaco12,sbordone10,lind09,mucciarelli12}. Regarding the
consistency of the data, we have followed the treatment of Ref.
\cite{pdgbook}.

\subsection{Results with $C=0$}

As a first case, we have computed the primordial abundance of light
nuclei for different values for the active-sterile neutrino mixing
at a fixed baryon density and a null initial condition for the
sterile neutrino (meaning $\zeta =0$). In order to obtain the best
values for the parameters, $\sin^2 2 \phi_1$ and $\sin^2 2 \phi_2$,
we have performed a $\chi^2$ test. Results are presented in the
first two rows of Table \ref{tabla1}.

\begin{table}[t!]
\begin{center}
\begin{tabular}{|c|c|c|c|c|}
\hline
Data & $\zeta \pm \sigma$ &$\sin^2 2\phi_1
\pm \sigma$  &$\sin^2 2\phi_2 \pm \sigma$&$
\frac{\chi^2}{N-a}$ \\ \hline
D+$^4$He+$^7$Li & fixed at 0 &$0.019^{+0.026}
$&$0.076^{+0.058}_{-0.055}$&$9.82$\\ \hline
D+$^4$He &fixed at 0 & $0.022^{+0.027}$&$0.093^{+0.063}_{-0.059}$&$1.04$\\\hline
D+$^4$He+$^7$Li & $0.30 \pm 0.12$
&$0.025^{+0.012}$&$0.000^{+0.093}$&$10.04$  \\ \hline
\end{tabular}
\caption{Best-fit parameter values and $1\sigma$ errors (case $C=0$).}
\label{tabla1}
\end{center}
\end{table}

The $\chi^2$-test indicates that the global fit is not a good one if
the complete set of data is included in the analysis, however, a
better fit is obtained if the data on primordial lithium are removed
from the sample of observational data. The best values for the
mixing angles are small in both cases and consistent with zero at
$2\sigma$-level.

Next, we have considered the parameter $\zeta$ as an extra parameter
to adjust. The results of the statistical analysis is presented in
the third row of Table \ref{tabla1}. Once again, the statistical
analysis does not give a good fit, the values for the mixing angles
remain small and consistent with a null mixing angle, meanwhile the
parameter $\zeta$ increases its value.

\subsection{Results with $C \ne 0$}

The next step in the analysis was to turn-on the interaction between
neutrinos and anti-neutrinos by setting $C \neq 0$. Once again, we
have computed neutrino occupation factors and primordial abundances
and compared them to the observational data through a statistical
test. The results, with a fixed value of the parameter $\zeta$ are
shown in the first two rows of the Table \ref{tabla2}.

\begin{table}[t!]
\begin{center}
\begin{tabular}{|c|c|c|c|c|}
\hline Data & $\zeta \pm \sigma$ &$\sin^2 2\phi_1 \pm \sigma$
&$\sin^2 2\phi_2 \pm \sigma$&$\frac{\chi^2}{N-a}$ \\ \hline
D+$^4$He+$^7$Li &  fixed at 0 &$0.002^{+0.007}$&$0.040 \pm
0.033$&$9.82$ \\ \hline D+$^4$He &  fixed at 0 &
$0.007^{+0.004}$&$0.020^{+0.078}_{-0.009}$&$1.04$\\ \hline
D+$^4$He+$^7$Li &
$0.30^{+0.11}_{-0.13}$&$0.000^{+0.004}$&$0.014^{+0.023}$&$10.04$  \\
\hline
\end{tabular}
\caption{Best-fit parameter values and $1\sigma$ errors considering
$C \ne 0$.} \label{tabla2}
\end{center}
\end{table}

Also for this case, there is not a good fit when all data are used
in the analysis, but it improves if one removes the data on $^7$Li.
The mixing angles are smaller that the ones obtained in the previous
section (case $C=0$).

As a final analysis, we have adjusted the renormalization parameter
$\zeta$. The fit is not a good one, however, the value for $\zeta$
is in agreement with the one obtained previously, and the mixing
angles remain small and consistent with a null value.

\section{Conclusion}
\label{conclusion}

In this work, we have included a sterile neutrino in the formalism
of primordial nucleo-synthesis and computed the abundances of the
light nuclei as a function of the mixing parameters. We have
computed neutrino occupation factors and neutron-to-proton decay
rates, as functions of the introduced mixing parameters, and
extracted their values from the comparison between calculated and
observed primordial abundances. We have found that the two  added
active-sterile mixing angles are small and that they are
consistent with zero at $1 \sigma$ or $2 \sigma$. The value
of the parameter $\zeta$, which is the renormalization factor of the
sterile neutrino thermal occupation, is found to be the same for the
two cases considered in this work (neutrino-antineutrino density
coupling $C=0$ and $C\ne 0$). The results are found to be consistent
with previous works
\cite{kishimoto06,civitarese08a,civitarese08b,mosquera11,mosquera14}.

\section*{{\bf Acknowledgments}}

Support for this work was provided by the National Research Council
(CONICET) of Argentina, and by the ANPCYT of Argentina. O. C. and M.
E. M. are members of the Scientific Research Career of the CONICET.

\end{document}